\documentclass[useAMS,usenatbib]{mn2e}
\usepackage{graphicx}
\def\aap{A\&A}

\def\apj{ApJ}
\def\apjs{ApJS}

\def\mnras{MNRAS}

\def\aapr{A\&ARev}

\def\gca{Geochimica et Cosmochimica Acta}

\title[X-ray study of PACWBs in NGC\,6604]{Long-term XMM-Newton investigation of two particle-accelerating colliding-wind binaries in NGC\,6604: HD\,168112 and HD\,167971.\thanks{Based on observations with XMM-Newton, an ESA Science Mission with instruments and contributions directly funded by ESA Member states and the USA (NASA).}}
\author[M. De Becker]{M. De Becker$^{1}$\thanks{E-mail:
debecker@astro.ulg.ac.be}\\
$^{1}$Department of Astrophysics, Geophysics and Oceanography, University of Li\`ege, Quartier Agora, 19c All\'ee du 6 Ao\^ut, \\ B5c, B-4000 Sart Tilman, Belgium\\
}

\begin{document}

\date{Accepted . Received ; in original form }

\pagerange{\pageref{firstpage}--\pageref{lastpage}} \pubyear{2012}

\maketitle

\label{firstpage}

\begin{abstract}
{ The long-term (over more than one decade) X-ray emission from two massive stellar systems known to be particle accelerators is investigated using XMM-Newton. Their X-ray properties are interpreted taking into account recent information about their multiplicity and orbital parameters. The two targets, HD\,168112 and HD\,167971 appear to be overluminous in X-rays, lending additional support to the idea that a significant contribution of the X-ray emission comes from colliding-wind regions. The variability of the X-ray flux from HD\,168112 is interpreted in terms of varying separation expected to follow the 1/D rule for adiabatic shocked winds. For HD\,167971, marginal decrease of the X-ray flux in September 2002 could tentatively be explained by a partial wind eclipse in the close pair. No long-term variability could be demonstrated despite the significant difference of separation between 2002 and 2014. This suggests the colliding-wind region in the wide orbit does not contribute a lot to the total X-ray emission, with a main contribution coming from the radiative shocked winds in the eclipsing pair. The later result provides evidence that shocks in a colliding-wind region may be efficient particle accelerators even in the absence of bright X-ray emission, suggesting particle acceleration may operate in a wide range of conditions. Finally, in hierarchical triple O-type systems, thermal X-rays do not necessarily constitute an efficient tracer to detect the wind-wind interaction in the long period orbit.}
\end{abstract}


\section{Introduction}
The open cluster NGC\,6604 is located is the Ser\,OB2 association, at a distance of 1.7\,kpc \citep{reipurth2008}. It harbours two members of the category of particle-accelerating colliding-wind binaries (PACWB), i.e. HD\,168112 and HD\,167971. This category of objects includes binary, or higher multiplicity systems made of massive stars known to be able to accelerate particles up to relativistic energies. The most up to date catalogue of such objects has been published by \citet{pacwbcata}. Even though PACWBs are mainly identified through radio observations revealing synchrotron radiation, these objects are potential non-thermal emitters in the high energy domain. Relativistic electrons -- accelerated via the Diffusive Shock Acceleration (DSA) process \citep{EU1993,Reitberger2014} -- can indeed be responsible for the production of high energy photons through inverse Compton scattering, with resulting energies depending on the frequency of the seed photons and on the energy of the scattering electrons \citep{BG1970}. However, it has been demonstrated that the detection of such a non-thermal soft X-ray emission component is very unlikely because of the strength of the thermal emission due to the colliding-winds \citep{debeckerreview}. Non-thermal X-rays are thus only expected to be detected in the hard X-ray domain (above 10\,keV), as demonstrated in the cases of $\eta$\,Car \citep{etacarnt} and WR\,140 \citep{wr140nt}.

\begin{figure*}
\begin{center}
\includegraphics[width=16cm]{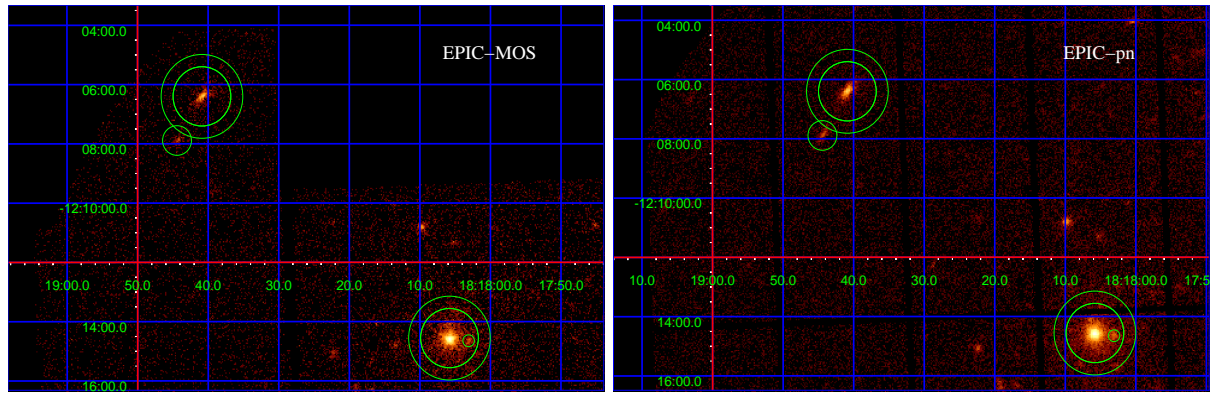}
\caption{EPIC-MOS1 (left) and EPIC-pn (right) images of NGC\,6604 between 0.3 and 10.0\,keV. The positions of the two targets are shown, along with the spatial filters used for the extraction of source and background spectra. HD\,168112 is located in the upper left part, and HD\,167971 is the brightest source in the lower right part. Only the September 2014 images are presented. We refer to \citet{DeBecker168112,DeBecker167971} for images of the previous epochs. North is up and the East is on the left.\label{epic}}
\end{center}
\end{figure*}

\begin{table*}
\caption{Journal of observations of NGC\,6604. \label{journal}}
\begin{center}
\begin{tabular}{l c c c c c c c c c}
\hline
Observ. Id. & Date & ID & \multicolumn{2}{c}{Performed duration (s)} & & \multicolumn{2}{c}{Eff. exp. time (s)} & Soft proton flare\\
\cline{4-5}\cline{7-8}
 &  &  & MOS & pn & & MOS & pn & \\
 (1) & (2) & (3) & (4) & (5) & & (6) & (7) & (8) \\
\hline
\vspace*{-0.2cm}\\
0008820301 & 7 April 2002 & A02 & 13120 & 10550 & & 9560 & 6820 & yes\\
0008820601 & 9 September 2002 & S02 & 13670 & 12050 & & 13400 & 10700 & no\\
0740990101 & 9 September 2014 & S14 & 25500 & 23840 & & 24300 & 19000 & yes\\
\vspace*{-0.2cm}\\
\hline
\end{tabular}
\end{center}
\end{table*}

Colliding-wind binaries are indeed well-known thermal X-ray emitters \citep{SBP1992,PP2010}. Their X-ray emission comes from the addition of the thermal emission from the stellar winds of individual stars with that of the colliding-wind region (CWR). Individual winds emit X-rays because of intrinsic shocks resulting from the line-driving instability, leading to slabs of wind material characterized by temperatures of a few 10$^6$\,K  \citep{FeldX}. On the other hand, the strong shocks due to colliding winds, with pre-shock velocities going up to 2000-3000\,km\,s$^{-1}$, are able to rise post-shock plasma temperatures up to several 10$^7$\,K \citep{SBP1992,PP2010}. Both systems investigated in this paper, HD\,168112 and HD\,167971, have been studied in X-rays using XMM-Newton more than a decade ago \citep{DeBecker168112,DeBecker167971}. Considering their multiple nature with rather long orbital time-scales, a longer term investigation was thought relevant. For this reason, they were observed again about 12 years later to investigate long-term trends of their X-ray properties.

The objective of this study is to complete the description of the X-ray behaviour of HD\,168112 and HD\,167971 with respect to previous studies, in light of new results published recently about their multiplicity \citep{hd167971vlti,Ibanoglu2013,Sana2014}. As well-studied PACWBs, they constitute potential targets for future studies aiming at modelling the particle acceleration and non-thermal emission processes at work in colliding-wind binaries. In this context, X-ray observations constitute a relevant probe of the properties of the shocks both responsible for their thermal X-ray emission and particle acceleration processes. The observations and data processing are described in Section\,\ref{obser}. The results obtained for both targets are presented and discussed in Sections\,\ref{168112} and \ref{167971}. The summary and conclusions are finally given in Section\,\ref{concl}. 

\section{Observations and data processing}\label{obser}

NGC\,6604 has been a target for XMM-Newton at three different dates, two in 2002 and one in 2014. The journal of observations is presented Table\,\ref{journal}. The observation identifier (col. 1) and date (col. 2) are provided, along with an abbreviated identifier (col 3) that will be used throughout the paper to distinguish the three epochs. Observations A02 and S14 were affected by soft proton flares (col. 8), justifying the significant difference between the performed duration (col. 4 and 5) and the effective exposure times (col. 6 and 7) of the observations. EPIC instruments \citep{mos,pn} were operated in Full Frame mode, and the medium filter was used. Data were processed using the XMM-Newton Science Analysis Software (SAS) v.14.0.0 on the basis of the Observation Data Files (ODF) provided by the European Space Agency (ESA). Event lists were filtered using standard screening criteria (pattern $\leq$\,12 for MOS and pattern $\leq$\,4 for pn). The data processing was performed using version 3.232 of the Calibration Database (release of 9 December 2014). Concerning the exposures affected by soft proton flares, light curves based on events with Pulse Invariant energy channels higher than 10000 were extracted \citep{Lumb2002}. For the April 2002 observation, time intervals with background level higher than 25 counts for MOS, and 150 counts for pn, were rejected. In September 2014, the rejection thresholds were 20 counts for MOS and 150 counts for pn data sets. We note that for the S14 observation, the aim point was set on the position of HD\,167971 but the signal in RGS spectra was too low to provide data of sufficient quality to warrant adequate spectral analysis. This study will therefore focus on EPIC data.

Our two targets appear in the field of view of the three observations, on different locations on the detector. In Figure\,\ref{epic}, HD\,168112 is located in the upper left part, and HD\,167971 appears in the lower right part. The events related to the source and background were extracted from the complete event list, without applying any energy band filter at this stage. 

The source extraction regions for both targets were circles with 60 arcsec radius, and background spectra were extracted in annular regions centered on the position of the source (with inner radius of 60 arcsec, and outer radius of 85 arcsec to obtain a background area very similar to that of the source). In the case HD\,167971, a 12 arcsec radius circular region was used to reject a point source located at coordinates [18:18:03.089,-12:14:38.90]. For HD\,168112, a 30 arcsec circular region was used to reject a point source at coordinates [18:18:44.414,-12:07:53.31] from the background region. These regions are shown in Fig.\,\ref{epic} for the EPIC-MOS1 and EPIC-pn data set, in September 2014. It must however be noted that for the EPIC-pn data in September 2014, different background regions were checked. Considering the position of the source close to the field of view boundary, the spectrum is contaminated by a fluorescence line at about 8\,keV, part of the so-called internal quiescent background (see the XMM-Newton User's Handbook and \citealt{Lumb2002}). Beside the 'classical' annular region described above, different circular regions were tested to try to get rid of the fluorescent feature, but the instrumental feature was still present. It is indeed very difficult to make those fluorescent features disappear. It must however be noted that the presence of that feature -- albeit visible in the spectrum (see lower panel of Fig.\,\ref{epic168112} -- is not significant in the context of the spectral analysis described in Section\,\ref{analysis1}. The spectral modelling is especially influenced by the real spectral features observed at lower energies, and contributing much more significantly to the EPIC spectrum.

The count rates obtained in the three EPIC data sets for both targets are quoted in Table\,\ref{epiccr}. Please note that the count rates do not translate the actual physical fluxes of the sources. These numbers are substantially affected by the position on the detector due to an non-homogeneous response. In particular, sources located at the center are more exposed than sources closer to the boundary of the field. This has to be taken into account while discussing the variability of sources located at different detector positions depending on epoch. For this reason, the discussion about the long term behaviour (over about a decade) of the targets in X-rays will be based on the results of the spectral modelling (see Sections\,\ref{var1} and \ref{var2}). In addition, one has to be especially careful for sources located close to the boundaries of the field of view. For instance, it is clear in Figure\,\ref{epic} that HD\,168112 is not optimally exposed in September 2014 for MOS instruments. This leads to a rather weak signal in the harder part of the EPIC bandpass where the source is less bright, as shown in spectra presented in the next sections. 

\begin{table}
\caption{Count rates measured by the three EPIC instruments (between 0.3 and 10\,keV) for the two targets at the three observation epochs. \label{epiccr}}
\begin{center}
\begin{tabular}{l c c}
\hline
 & HD\,168112 & HD\,167971 \\
\hline
\vspace*{-0.2cm}\\
\multicolumn{3}{c}{April 2002}\\
\hline
\vspace*{-0.2cm}\\
MOS1 & 0.059$\pm$0.003 cnt\,s$^{-1}$ & 0.148$\pm$0.004 cnt\,s$^{-1}$ \\
MOS2 & 0.055$\pm$0.003 cnt\,s$^{-1}$ & 0.144$\pm$0.004 cnt\,s$^{-1}$ \\
pn & 0.173$\pm$0.007 cnt\,s$^{-1}$ & 0.401$\pm$0.010 cnt\,s$^{-1}$ \\
\hline
\vspace*{-0.2cm}\\
\multicolumn{3}{c}{September 2002}\\
\hline
\vspace*{-0.2cm}\\
MOS1 & 0.036$\pm$0.002 cnt\,s$^{-1}$ & 0.138$\pm$0.004 cnt\,s$^{-1}$ \\
MOS2 & 0.041$\pm$0.002 cnt\,s$^{-1}$ & 0.160$\pm$0.004 cnt\,s$^{-1}$ \\
pn & 0.126$\pm$0.005 cnt\,s$^{-1}$ & 0.476$\pm$0.008 cnt\,s$^{-1}$ \\
\hline
\vspace*{-0.2cm}\\
\multicolumn{3}{c}{September 2014}\\
\hline
\vspace*{-0.2cm}\\
MOS1 & 0.034$\pm$0.002 cnt\,s$^{-1}$ & 0.245$\pm$0.004 cnt\,s$^{-1}$ \\
MOS2 & 0.035$\pm$0.002 cnt\,s$^{-1}$ & 0.235$\pm$0.003 cnt\,s$^{-1}$ \\
pn & 0.106$\pm$0.003 cnt\,s$^{-1}$ & 0.834$\pm$0.008 cnt\,s$^{-1}$ \\
\vspace*{-0.2cm}\\
\hline
\end{tabular}
\end{center}
\end{table}

For all spectra at every epoch, response matrices and ancillary response files were computed using the {\tt rmfgen} and {\tt arfgen} metatasks. Spectra were grouped to get at least 20 events per energy bin. The spectral analysis was performed using the XSPEC software (v.12.8.2, \citealt{xspec1996,xspec2001}). Spectral fitting was executed using the $\chi^2_\nu$ value as goodness-of-fit indicator, which is adequate as the grouping criterion leads to a sufficient number of counts in each energy bin to assume a Gaussian distribution. All error bars were computed on the basis of the {\tt error} command within XSPEC. This leads to an adequate estimate of confidence intervals of model parameters based on a systematic exploration of the parameter space. Such a procedure allows to refine the spectral fitting and the convergence to a best-fit model. Details on the spectral analysis for HD\,168112 and HD\,167971 are provided in Sections \ref{analysis1} and \ref{analysis2}, respectively.

\section{HD\,168112}\label{168112}

\subsection{The system}\label{system1}
HD\,168112 (BD\,--12 4988) includes an O5.5III(f$^+$) primary with an unidentified secondary. This star is known as a non-thermal radio emitter since the large campaigns in the 1980's dedicated to massive stars \citep{BAC1989}, with a well-established negative spectral index pointing to a clear synchrotron signature. In the context of the so-called 'standard model' for particle acceleration, this object was suspected to be a binary with a non-thermal emission region coincident with a colliding-wind region. However, the search for a spectroscopic companion failed to detect any \citep{rauw2005}. The solution came from another technique with the development of long baseline interferometry which allowed to resolve a companion in HD\,168112 using the Very Large Telescope Interferometer (VLTI, Chile). \citet{Sana2014} reported on the presence of a companion with a separation of a 3.33$\pm$0.17\,mas\ in May 2012, providing evidence for the binary nature of this object. At a distance of 1.7\,kpc, this separation translates into a projected linear separation of the order of 6\,au. The interferometric study revealed a magnitude difference $\Delta$H = 0.17 $\pm$ 0.19. On the basis of the synthetic photometry published by \citet{MP2006}, this confidence interval of $\Delta$H suggests spectral classifications for the secondary ranging between O5.5III and O7.5III. One extreme value of $\Delta$H is also compatible with an O4V spectral type for the secondary, but such a classification would have been revealed by the existing visible spectra of the system, but this is not the case. 

Before the advent of VLTI results, additional hints for the binary nature came from XMM-Newton observation revealing a spectrum too hard to be compatible with that of a single O-type star, along with a weak variation of the X-ray flux between two epochs separated by 5 months \citep{DeBecker168112}. Such an intrinsic variation from an individual stellar wind is indeed very unlikely. On the other hand, radio observations revealed also a significant variability of HD\,168112 \citep{DeBecker168112,Blomme168112}. On the basis of the radio light curve exploiting data spread over about 20 years, \citet{Blomme168112} explored potential orbital time-scales but the time series was too sparse to be conclusive. So far, the orbital period is still undetermined.

When one is dealing with colliding-winds in massive binaries, it is relevant to introduce the cooling parameter defined by \citet{SBP1992}. This parameter is basically a measurement of the ratio of the cooling time of the shocked gas over the flow time. A value lower or close to one points to a radiative regime where the X-ray emission is typically a fraction of the kinetic power injected in the wind collision (whatever the stellar separation). Alternatively, a cooling parameter significantly larger than one characterizes an adiabatic regime, with an X-ray emission from the shocked gas inversely proportional to the stellar separation (the so-called 1/D-law, where D is the separation). Considering the typical values for the mass loss rate and terminal velocity of the O5.5III primary, one gets $\mathrm{{\dot M} = 1.35\,\times\,10^{-6}\,M_\odot\,yr^{-1}}$, and V$_\infty$ = 2800\,km\,s$^{-1}$ \citep{Muijres2012}. The definition of the cooling parameter given by \citet{SBP1992} can be translated into the following criterion that gives a lower limit on the distance (in cm) between a star and the stagnation point warranting an adiabatic regime:
$$\mathrm{d_{min} = 10^{31}\,\frac{\dot M}{v_\infty^4}}$$
with quantities expressed in the same units as above. This translates into a minimum distance of about 0.015\,au. Even considering that in a highly eccentric orbit the winds may collide at a very short distance from the star with the weaker wind at periastron, $\mathrm{d_{min}}$ is much smaller than the projected separation measured with the VLTI, suggesting an adiabatic regime. One is lacking the required accurate information to apply this criterion to the secondary object. However, assuming the spectral type range mentioned above in this Section, the wind parameters should be similar to that of the  the primary or take the following values if the spectral type is O7.5III: $\mathrm{{\dot M} = 2.5\,\times\,10^{-7}\,M_\odot\,yr^{-1}}$ and V$_\infty$ = 4400\,km\,s$^{-1}$ \citep{Muijres2012}.  The latter assumption leads to a lower limit on the distance between the colliding-wind region and the secondary of about 0.0004\,au for the adiabatic regime (much smaller than the projected separation measured by \citealt{Sana2014}). One may therefore reasonably consider that the expected properties of the shocked winds in HD\,168112 point to the adiabatic regime even in a highly eccentric system. 

\subsection{X-ray properties}\label{analysis1}
It is worth to point to some differences with respect to the analysis performed by \citet{DeBecker168112} on the basis of the A02 and S02 data sets. Notably, the event lists were filtered to get rid of time intervals contaminated by a soft proton flare. This leads the A02 exposure to be reduced by about 30\,\% with respect to previous analysis (see Table\,\ref{journal}), but the spectra are expected to be cleaner.

\begin{table*}
\caption{Best-fit parameters for HD\,168112 using the {\tt wabs*wind*(apec+apec)} model (in the 0.3--10\,keV energy band). The error bars represent the 90\,\% confidence level.\label{fit168112TT}}
\begin{center}
\begin{tabular}{l c c c c c c c}
\hline
Instr. & N$_{wind}$ & kT$_1$ & N$_1$ & kT$_2$ & N$_2$ & $\chi^2_\nu$ (d.o.f.) & f$_{obs}$ \\
  & (10$^{22}$\,cm$^{-2}$) & (keV) & (10$^{-3}$\,cm$^{-5}$) & (keV) & (10$^{-3}$\,cm$^{-5}$) &  & (erg\,cm$^{-2}$\,s$^{-1}$) \\
\hline
\vspace*{-0.2cm}\\
\multicolumn{8}{c}{April 2002}\\
\hline
\vspace*{-0.2cm}\\
MOS & 0.71$_{-0.20}^{+0.18}$ & 0.29$_{-0.04}^{+0.06}$ & 5.50$_{-2.67}^{+5.70}$ & 3.68$_{-1.48}^{+5.86}$ & 0.31$_{-0.09}^{+0.10}$ & 1.10 (59) & 5.4$_{-2.3}^{+0.2}$\,$\times$\,10$^{-13}$ \\
\vspace*{-0.2cm}\\
pn & 0.81$_{-0.23}^{+0.17}$ & 0.28$_{-0.03}^{+0.06}$ & 8.13$_{-4.53}^{+6.15}$ & 2.47$_{-0.70}^{+1.83}$ & 0.53$_{-0.17}^{+0.14}$ & 1.08 (70) & 6.5$_{-3.1}^{+0.2}$\,$\times$\,10$^{-13}$ \\
\vspace*{-0.2cm}\\
EPIC & 0.75$_{-0.14}^{+0.14}$ & 0.29$_{-0.03}^{+0.03}$ & 6.55$_{-2.53}^{+4.10}$ & 3.34$_{-0.89}^{+2.12}$ & 0.34$_{-0.07}^{+0.08}$ & 1.15 (142) & 5.8$_{-0.7}^{+0.4}$\,$\times$\,10$^{-13}$ \\
\hline
\vspace*{-0.2cm}\\
\multicolumn{8}{c}{September 2002}\\
\hline
\vspace*{-0.2cm}\\
MOS & 0.53$_{-0.17}^{+0.17}$ & 0.30$_{-0.02}^{+0.04}$ & 2.60$_{-0.85}^{+1.30}$ & 2.54$_{-0.77}^{1.94}$ & 0.26$_{-0.07}^{+0.08}$ & 0.97 (55) & 3.8$_{-0.6}^{+0.3}$\,$\times$\,10$^{-13}$\\
\vspace*{-0.2cm}\\
pn & 0.81$_{-0.18}^{+0.31}$ & 0.27$_{-0.04}^{+0.02}$ & 7.53$_{-1.12}^{+12.00}$ & 2.95$_{-0.93}^{+2.06}$ & 0.38$_{-0.09}^{+0.12}$ & 0.77 (76) & 5.6$_{-2.2}^{+1.0}$\,$\times$\,10$^{-13}$\\
\vspace*{-0.2cm}\\
EPIC & 0.72$_{-0.13}^{+0.12}$ & 0.25$_{-0.02}^{+0.05}$ & 7.02$_{-3.82}^{+3.78}$ & 2.45$_{-0.62}^{+0.93}$ & 0.33$_{-0.06}^{+0.07}$ & 1.07 (149) & 4.4$_{-0.8}^{+0.3}$\,$\times$\,10$^{-13}$\\
\hline
\vspace*{-0.2cm}\\
\multicolumn{8}{c}{September 2014}\\
\hline
\vspace*{-0.2cm}\\
MOS & 0.42$_{-0.18}^{+0.19}$ & 0.32$_{-0.03}^{+0.03}$ & 2.53$_{-0.92}^{+1.26}$ & 2.11$_{-0.25}^{+0.52}$ & 0.72$_{-0.10}^{+0.07}$ & 1.02 (85) & 7.3$_{-0.9}^{+0.4}$\,$\times$\,10$^{-13}$\\
\vspace*{-0.2cm}\\
pn & 0.46$_{-0.16}^{+0.15}$ & 0.27$_{-0.02}^{+0.04}$ & 2.64$_{-1.17}^{+1.63}$ & 1.57$_{-0.13}^{+0.14}$ & 0.75$_{-0.08}^{+0.07}$ & 1.15 (115) & 5.9$_{-0.9}^{+0.2}$\,$\times$\,10$^{-13}$\\
\vspace*{-0.2cm}\\
EPIC & 0.46$_{-0.14}^{+0.12}$ & 0.31$_{-0.03}^{+0.04}$ & 2.53$_{-1.02}^{+1.39}$ & 1.90$_{-0.17}^{+0.19}$ & 0.72$_{-0.05}^{+0.06}$ & 1.15 (218) & 6.6$_{-0.7}^{+0.1}$\,$\times$\,10$^{-13}$\\
\vspace*{-0.2cm}\\
\hline
\end{tabular}
\end{center}
\end{table*}

Spectra were first fitted using two-temperature models of the type {\tt wabs*wind*(apec+apec)}. The {\tt wabs} component stands for the photoelectric absorption of X-rays by the interstellar material along the line of sight, and it is based on abundances by \citet{anders1982}. Using the same approach as described by \citet{DeBecker168112}, it was fixed to a value of 0.58\,$\times$\,10$^{22}$\,cm$^{-2}$. This value was obtained assuming $\mathrm{N_H}$ = 5.8\,$\times$\,10$^{21}$\,$\times$\,E(B -- V)\,cm$^{-2}$ \citep{Boh}, with a colour excess determined from an observed (B -- V) of +0.69 \citep{Chlebowski1989} and an intrinsic colour (B -- V)$_\circ$ of --0.309 \citep{MB1981}. The {\tt wind} component accounts for the absorption of X-rays by the wind (i.e. circumstellar) material. This absorption component was computed assuming the presence of ionized material made of the ten most abundant elements with solar abundances \citep{andersgrevesse}. Details on this ionized absorption model are given in \citet{DeBecker167971}, and references therein. The {\tt apec} emission components are used to reproduce the thermal X-ray emission spectrum from an optically thin plasma heated by shocks, either intrinsic or due to colliding-winds. A typical plasma temperature of about 3\,$\times$\,10$^{6}$\,K is obtained for the soft emission component. The temperature of the hard component is less well determined, with values in the range of 15-30\,$\times$\,10$^{6}$\,K. We note that in the approach followed by \citet{DeBecker168112} the fit were not optimized by an exploration of the parameter space using the {\tt error} command within XSPEC. The exploration of the best-fit parameter space revealed convergence difficulties which were not apparent at the time of the previous analysis. The problem comes from the existence of different local minima, in particular characterized by kT values of about 0.3 or 0.7\,keV for the soft component. This happens only for a few spectra in the series. Typically, for the lower quality spectra, a degeneracy may appear between models with different temperatures weighted by different normalization parameters. The kT value of 0.3\,keV appears to be consistently adequate for all spectra at every epoch. However, the highest quality spectra did not stabilize in solutions including a plasma temperature of about 0.7\,keV for the soft component. For these reasons, the solution with the lower value for kT$_1$ was privileged, in agreement with the results obtained by \citet{DeBecker168112}.

\begin{figure}
\begin{center}
\includegraphics[width=8cm]{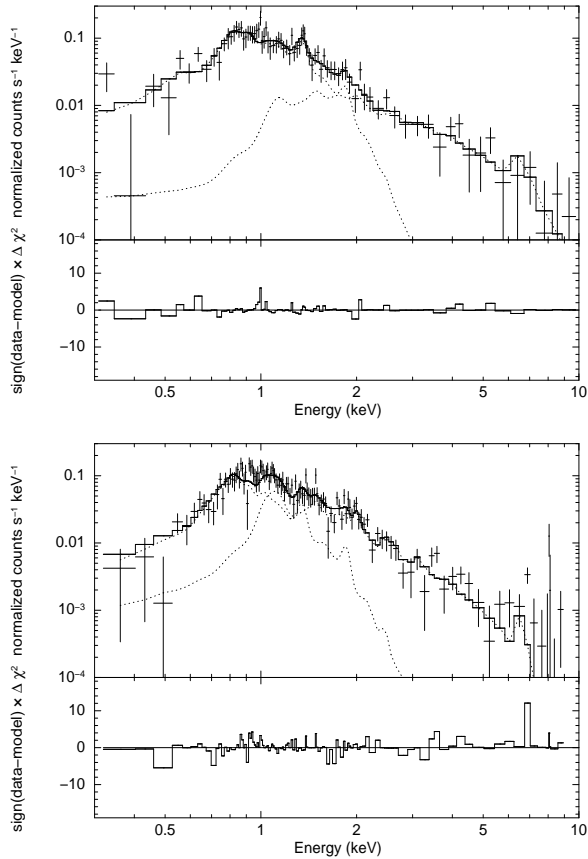}
\caption{Fit of the two-temperature model to EPIC-pn spectra obtained in September 2002 (top) and September 2014 (bottom) between 0.3 and 10.0\,keV. In both cases, the lower panel shows the residuals in the sense data minus model. The apparent spectral feature at about 8\,keV in September 2014 results from the fluorescence of the pn enclosure hit by charged particles.\label{epic168112}}
\end{center}
\end{figure}

Best-fit parameters are summarized in Table\,\ref{fit168112TT} for EPIC-MOS1 and EPIC-MOS2 data simultaneously, for EPIC-pn individually and for the combined EPIC instruments. Fluxes have also been determined along with error bars based on the exploration of the parameter confidence ranges of the best fit. Globally, the soft plasma temperature is constant throughout the time series. The hard temperature does not seem to present a clear trend as a function of time, as discrepancies appear between different data sets of a given epoch. The most plausible variations concern the wind absorption component and the normalization parameters of the emission components. On average, the faintest and brightest spectra are associated to the S02 and S14 epochs, respectively. EPIC spectra with the associated best-fit models are shown in Fig.\,\ref{epic168112}. The use of a three-temperature model did not improve the results of the spectral fitting. Typical reduced $\chi^2$ values were indeed slightly lower than those quoted in Table\,\ref{fit168112TT}. For instance, for combined EPIC instruments the values were 1.15, 0.98 and 1.03 respectively for the three epochs. However, the three emission components model led to kT values of about 0.2--0.3 and 0.8--1.0\,keV, with a very faint third component characterized by an excessive kT value (8 -- 15\,keV) with huge relative errors (often close to 100\%). This suggests clearly that the use of a third component does not provide any improvement of the physical interpretation of the present data set. Finally, as justified in \citet{debeckerreview}, the use of a power law component to account for a potential inverse Compton scattering process was not considered as one is dealing only with soft X-rays (below 10\,keV) dominated by thermal emission.

\subsection{Discussion}\label{disc1}

\subsubsection{The X-ray luminosity}\label{lumin1}
The overall X-ray emission from HD\,168112 corrected for interstellar absorption corresponds to a flux of 2.6--3.0\,$\times$\,10$^{-12}$\,erg\,cm$^{-2}$\,s$^{-1}$. Considering the distance to NGC\,6604 is 1700\,pc, this translates into a luminosity (L$_X$) range of 0.9--1.0\,$\times$\,10$^{33}$\,erg\,s$^{-1}$. On the basis of the calibration of parameters given by \citet{Martins2005}, the bolometric luminosity of the primary should be 1.7\,$\times$\,10$^{39}$\,erg\,s$^{-1}$. As the spectral classification of the secondary is not determined, one can only speculate on its main properties. On the basis of the discussion in Section\,\ref{system1}, a likely range for the secondary bolometric luminosity is 0.9--1.7\,$\times$\,10$^{39}$\,erg\,s$^{-1}$. These values translate into a L$_X$/L$_{bol}$ ratio of about 2.6--3.8\,$\times$\,10$^{-7}$, pointing a to a significant -- although weak -- X-ray overluminosity, with respect to the L$_X$/L$_{bol}$ ratio of $\sim$\,10$^{-7}$ expected for single O-star winds \citep{lxlbol1989,lxlbol1991,owocki2013}. This results demonstrates that the colliding-wind region in the HD\,168112 system is moderately bright in X-rays. This is in agreement with the global shape of the EPIC spectrum which presents a noticeable -- albeit not spectacular -- hardness.

The soft X-ray emission (between 0.3 and 10.0 keV) from the colliding winds could be roughly estimated, subtracting the expected contributions from individual stars. Assuming a 10$^{-7}$ luminosity ratio for the stars in the system, we obtain 1.7\,$\times$\,10$^{32}$\,erg\,s$^{-1}$ for the primary and 0.9--1.7\,$\times$\,10$^{32}$\,erg\,s$^{-1}$ for the secondary. These numbers lead to an X-ray luminosity for the colliding-wind region of about 6.5\,$\times$\,10$^{32}$\,erg\,s$^{-1}$.

The latter number could be compared to the kinetic power of the stellar winds: 
$$P_{kin} = \frac{1}{2}\,{\dot M}\,V_\infty^2$$
This quantity can be calculated using the typical stellar wind parameters published by \citet{Muijres2012}, as used in Section\,\ref{system1}. One obtains a total (for both stars) kinetic power in the range of 4.8--6.6\,$\times$\,10$^{36}$\,erg\,s$^{-1}$, translating into a L$_X$/P$_{kin}$ ratio of 0.9--1.5\,$\times$\,10$^{-4}$. These numbers are in agreement with the expected order of magnitude mentioned by \citep{pacwbcata}. Such a ratio can not be used as a immediate proxy for the power injection into the colliding-winds, as it will depend notably on the geometry of colliding-wind region which is not determined in this poorly characterized system. However, the expected future determination of the orbital parameters of HD\,168112 along with ongoing theoretical developments should allow us to address with more details the issue of the energy budget in this system. The latter point is crucial the discuss the energy injection rate into non-thermal processes, and in particular in the particle acceleration process which takes place in this system.

\subsubsection{Variability}\label{var1}
Beside the main spectral properties (essentially in terms of plasma temperature) found in agreement with the results published by \citet{DeBecker168112}, the results presented in Section\,\ref{analysis1} deserve to be discussed from the variability point of view. In the case of a binary system, the most straightforward source of variability could be attributed to the orbital motion. In an eccentric orbit, the phase-dependent separation will lead to changes in the physical properties of the colliding-wind region where a significant fraction of the X-ray emission is produced. As explained in Section\,\ref{system1}, in the adiabatic regime the intrinsic X-ray emission (corrected for both interstellar and wind absorption) is expected to be inversely proportional to the stellar separation. On the other hand, the absorbing column made of wind material is expected to change with the orbital phase. This orientation effect will also induce variations in the observed spectrum. However, provided that pre-shock velocities of the colliding winds do not change, one should not expect a change in the plasma temperatures from one epoch to the other. In an eccentric orbit, this must be true provided the winds have reached their terminal velocities before colliding at all observed orbital phases. 

As a consequence, the clearer trends are provided by the wind absorption component and the normalization (related to the intrinsic X-ray luminosity) of the emission component(s). The former point is clearly an orientation effect which depends intimately on the geometry of the system as seen by the observer. Without any indication on the orbital parameters of HD\,168112, the interpretation of these trends in terms of geometrical consideration would be highly speculative.

Concerning the varying intrinsic X-ray luminosity, the values can be estimated on the basis the best-fit parameters for EPIC instruments found in Table\,\ref{fit168112TT}. Intrinsic X-ray fluxes are 1.49\,$\times$\,10$^{-11}$, 1.51\,$\times$\,10$^{-11}$ and  0.68\,$\times$\,10$^{-11}$\,erg\,cm$^{-2}$\,s$^{-1}$ for the A02, S02 and S14 epochs. For a distance of about 1700\,pc, these fluxes translate into X-ray luminosities of 5.2\,$\times$\,10$^{33}$, 5.2\,$\times$\,10$^{33}$ and 2.4\,$\times$\,10$^{33}$\,erg\,s$^{-1}$. In the context of adiabatic shocked winds this quantity should follow a 1/D trend. It is therefore reassuring to see that the two 2002 observations give an almost identical value of the intrinsic X-ray luminosity, and the significantly different value for 2014 points very likely to a significantly different separation between the two stars at these two epochs. 

The eccentricity (e) of the orbit is related to the separation ratio at apastron ($\mathrm{D_A}$) and periastron ($\mathrm{D_P}$) by the following identify,
$$\mathrm{\Big(\frac{D_P}{D_A}\Big)_r = \frac{1 - e}{1 + e}}$$
where the index r stands for 'real', as opposed to m for 'measured' (see below).

On the other hand, on the basis of the 1/D-law, one can write that the measured separation ratio is
$$\mathrm{\Big(\frac{D_{14}}{D_{02}}\Big)_m = \frac{L_{X,02}}{L_{X,14}} = \xi = 2.2}$$
where $\mathrm{L_{X,14}}$ and $\mathrm{L_{X,02}}$ are the minimum and maximum measured intrinsic X-ray luminosities. 

Considering that 
$$\mathrm{\Big(\frac{D_P}{D_A}\Big)_r \leq \frac{1}{\xi}}$$
we have 
$$\mathrm{e \geq \frac{\xi - 1}{\xi + 1}}$$
With the $\xi$ value estimated above, one obtains a rough lower limit on the eccentricity equal to 0.38.

One should also mention that the measured X-ray emission does not come only from the colliding-wind region, but individual winds contribute also. The variable part of the overall emission appears thus on top of a constant emission. It means that the $\xi$ value constitutes itself a lower limit on the flux ratio coming from the colliding-wind region alone. The above lower limit on the eccentricity is therefore highly conservative.

\begin{table*}
\caption{Best-fit parameters for HD\,167971 using the {\tt wabs*wind*(apec+apec)} model (in the 0.3--10\,keV energy band). The error bars represent the 90\,\% confidence level.\label{fit167971TT}}
\begin{center}
\begin{tabular}{l c c c c c c c}
\hline
Instr. & N$_{wind}$ & kT$_1$ & N$_1$ & kT$_2$ & N$_2$ & $\chi^2_\nu$ (d.o.f.) & f$_{obs}$ \\
  & (10$^{22}$\,cm$^{-2}$) & (keV) & (10$^{-2}$\,cm$^{-5}$) & (keV) & (10$^{-3}$\,cm$^{-5}$) &  & (erg\,cm$^{-2}$\,s$^{-1}$) \\
\hline
\vspace*{-0.2cm}\\
\multicolumn{8}{c}{April 2002}\\
\hline
\vspace*{-0.2cm}\\
MOS & 0.58$_{-0.10}^{+0.11}$ & 0.28$_{-0.04}^{+0.04}$ & 1.33$_{-0.54}^{+1.07}$ & 0.99$_{-0.07}^{+0.10}$ & 2.43$_{-0.39}^{+0.41}$ & 1.14 (121) & 1.61$_{-0.16}^{+0.03}$\,$\times$\,10$^{-12}$ \\
\vspace*{-0.2cm}\\
pn & 0.51$_{-0.12}^{+0.11}$ & 0.27$_{-0.03}^{+0.03}$ & 1.34$_{-0.51}^{+0.77}$ & 1.16$_{-0.22}^{+0.10}$ & 2.27$_{-0.28}^{+0.78}$ & 1.09 (129) & 1.68$_{-0.10}^{+0.07}$\,$\times$\,10$^{-12}$ \\
\vspace*{-0.2cm}\\
EPIC & 0.56$_{-0.07}^{+0.07}$ & 0.26$_{-0.02}^{+0.03}$ & 1.41$_{-0.48}^{+0.56}$ & 1.00$_{-0.05}^{+0.05}$ & 2.57$_{-0.31}^{+0.27}$ & 1.16 (266) & 1.62$_{-0.06}^{+0.03}$\,$\times$\,10$^{-12}$ \\
\hline
\vspace*{-0.2cm}\\
\multicolumn{8}{c}{September 2002}\\
\hline
\vspace*{-0.2cm}\\
MOS & 0.49$_{-0.09}^{+0.09}$ & 0.30$_{-0.03}^{+0.03}$ & 0.92$_{-0.30}^{+0.39}$ & 1.20$_{-0.07}^{+0.08}$ & 1.74$_{-0.19}^{+0.20}$ & 1.10 (162) & 1.43$_{-0.10}^{+0.02}$\,$\times$\,10$^{-12}$ \\
\vspace*{-0.2cm}\\
pn & 0.47$_{-0.08}^{+0.08}$ & 0.25$_{-0.02}^{+0.02}$ & 1.17$_{-0.34}^{+0.44}$ & 0.98$_{-0.05}^{+0.07}$ & 1.98$_{-0.13}^{+0.20}$ & 0.93 (220) & 1.34$_{-0.04}^{+0.03}$\,$\times$\,10$^{-12}$ \\
\vspace*{-0.2cm}\\
EPIC & 0.49$_{-0.07}^{+0.06}$ & 0.28$_{-0.02}^{+0.01}$ & 1.09$_{-0.23}^{+0.35}$ & 1.12$_{-0.15}^{+0.09}$ & 1.70$_{-0.13}^{+0.12}$ & 1.09 (395) & 1.39$_{-0.04}^{+0.03}$\,$\times$\,10$^{-12}$ \\
\hline
\vspace*{-0.2cm}\\
\multicolumn{8}{c}{September 2014}\\
\hline
\vspace*{-0.2cm}\\
MOS & 0.52$_{-0.05}^{+0.05}$ & 0.27$_{-0.01}^{+0.02}$ & 1.35$_{-0.23}^{+0.28}$ & 1.18$_{-0.04}^{+0.04}$ & 2.00$_{-0.11}^{+0.10}$ & 1.41 (259) & 1.57$_{-0.03}^{+0.04}$\,$\times$\,10$^{-12}$ \\
\vspace*{-0.2cm}\\
pn & 0.53$_{-0.06}^{+0.04}$ & 0.28$_{-0.01}^{+0.01}$ & 1.32$_{-0.23}^{+0.24}$ & 1.22$_{-0.04}^{+0.04}$ & 2.05$_{-0.12}^{+0.10}$ & 1.08 (402) & 1.61$_{-0.01}^{+0.02}$\,$\times$\,10$^{-12}$ \\
\vspace*{-0.2cm}\\
EPIC & 0.52$_{-0.04}^{+0.04}$ & 0.28$_{-0.01}^{+0.01}$ & 1.27$_{-0.17}^{+0.17}$ & 1.20$_{-0.03}^{+0.03}$ & 2.01$_{-0.08}^{+0.08}$ & 1.38 (685) & 1.59$_{-0.02}^{+0.01}$\,$\times$\,10$^{-12}$ \\
\vspace*{-0.2cm}\\
\hline
\end{tabular}
\end{center}
\end{table*}

\section{HD\,167971}\label{167971}

\subsection{The system}\label{system2}

HD\,167971 (BD\,--12 4980, MY\,Ser) has been known as a potential triple system since the 1980's, with the studies by \citet{Leitherer1987} and \citet{DF1988}. It consists in a close eclipsing binary with a period of about 3.32 days, with a third object evolving on a much wider orbit. First evidence for the gravitational link between the close pair and the third star was provided by VLTI observations revealing a time-dependent astrometry suggestive of an orbital motion \citep{hd167971vlti}. More recently, \citet{Ibanoglu2013} published the very first orbital solution for the wide orbit. According to this spectroscopic solution the period should be 21.2$\pm$0.7\,yr, with an eccentricity of 0.53$\pm$0.05 and a projected semi-major axis of 5.5$\pm$0.7\,au. The spectral types are O7.5III and O9.5III for the close pair, and the spectral classification of the third component is O9.5-B0(I or III). 

\begin{figure}
\begin{center}
\includegraphics[width=7.5cm]{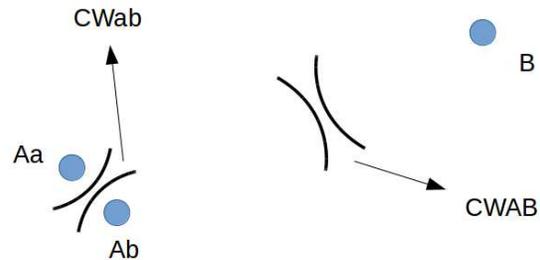}
\caption{Schematic view of the triple system HD\,167971. The thick black curves represent the shocks of the colliding-wind regions. Two colliding-wind regions are present: CWab between components Aa and Ab, and CWAB between components A and B. The relative stellar separations and sizes are not to scale.\label{triple}}
\end{center}
\end{figure}

The orbital period determined by \citet{Ibanoglu2013} is in excellent agreement with the variation time-scale suggested by the radio light curve published by \citet{Blomme167971}. This lends strong support to the idea that the non-thermal radio emission occurring in the wide orbit. In such a hierarchized system, one may distinguish two colliding-wind regions (see \citealt{hd167971vlti} for a discussion). The situation is illustrated in Fig.\,\ref{triple}. CWab and CWAB stand for the two expected colliding-wind regions, in the short and long period orbits respectively.

The analysis of the X-ray spectra at epochs A02 and S02 performed by \citet{DeBecker167971} revealed a significantly hard spectrum, with a warm component temperature of several 10$^{7}$\,K, in addition to a dominant softer emission. It suggested also a slight -- but marginal -- flux variation between the two epochs. The origin of that variation remained unclear. At the time of the \citet{DeBecker167971} study, no orbital parameter was known for the third object. It was not even clear that there was a gravitational relation between components A and B. Even though a possible scenario could be an orbital modulation of the X-ray emission from CWAB (see Figure\,\ref{triple}), a potential partial eclipse within the close Aab pair could not completely be rejected.

As far as colliding-wind regions are involved in the X-ray emission in HD\,167971, the issue of the regime of the shocked wind -- radiative or adiabatic -- is worth a discussion. Within the close Aab pair, the cooling parameter defined by \citet{SBP1992} can easily be estimated. The typical wind parameters for components Aa and Ab can be taken from \citet{Muijres2012}: $\mathrm{{\dot M}_{Aa} = 2.5\,\times\,10^{-7}\,M_\odot\,yr^{-1}}$, $\mathrm{{\dot M}_{Ab} = 1.4\,\times\,10^{-7}\,M_\odot\,yr^{-1}}$, $\mathrm{V_{\infty,Aa}}$ = 4400\,km\,s$^{-1}$ and $\mathrm{V_{\infty,Ab}}$ = 2600\,km\,s$^{-1}$. According to \citet{Ibanoglu2013}, the projected semi-major axis of the eclipsing binary (with a circular orbit) is about 36\,R$_\odot$. Such a short distance does certainly not allow the winds to reach their terminal velocities before colliding. The pre-shock velocities must therefore be estimated on the basis of the classical $\beta$-law for wind velocities, assuming $\beta$\,=\,0.8 for O-stars \citep{Puls2008}: i.e. $\mathrm{V_{Aa}}$ = 1300\,km\,s$^{-1}$ and $\mathrm{V_{Ab}}$ = 900\,km\,s$^{-1}$. These values lead to cooling parameter values of the order of unity, pointing to a radiative regime. Concerning CWAB involving the shocked wind of component B, the much larger separation leads inevitably to an adiabatic regime.

\subsection{X-ray properties}\label{analysis2}

EPIC spectra were analyzed according to the same approach as described in Section\,\ref{analysis1}. Good results were obtained using a two-temperature model made of the same components as for HD\,168112. For the insterstellar absorption, the same approach as described in Sect.\,\ref{analysis1} was followed. On the basis an observed (B -- V) of +0.77 \citep{Chlebowski1989} and an intrinsic colour (B -- V)$_\circ$ of --0.31 \citep{MB1981}, a N$_H$ value of 0.63\,$\times$\,10$^{22}$\,cm$^{-2}$ was determined and adopted for all models \citep[see][]{DeBecker167971}. The interstellar column was fixed to that value, allowing all other parameters to vary. The results are summarized in Table\,\ref{fit167971TT}.

The best-fit temperatures (kT) are about 0.3\,keV and 1.0-1.2\,keV, respectively for the soft and hard components. The consistency between results obtained with MOS and pn data is very good, except for the S02 epoch. The measured flux is very similar at all epochs, with only a slight decrease in S02. However, this trend is present only in MOS data, and not in the pn data set. It must be reminded that in the \citet{DeBecker167971} study, only MOS data were considered for epoch A02 and only pn data were considered for epoch S02. The reason could be the poor handling of gaps and bad columns be the first releases of the Science Analysis Software. Even though the data processing with the most recent releases of the SAS is expected to apply an adequate correction, the quality of the flux measurement may be questioned. For this reason, we will refrain to base our discussion on the pn result for the S02 epoch. 

The use of a three-temperature model did not improve the results on the complete time series. The typical plasma temperature of the soft and medium components were consistently of the order of 0.25-0.30\,keV and 0.95--1.00\,keV, respectively. However, the lack of significant signal in the higher energy part of the EPIC spectrum led to a difficulty to constrain both the temperature and the normalization parameter of the hardest component. The only relevant result with a three-temperature model was certainly obtained for the September 2014 data set, where HD\,167971 is well exposed at the center of the field of view. For this specific data set, some improvement was obtained by switching to a three-temperature model. The results are presented in Table\,\ref{3T167971}. It is notably noticeable in the plot shown in Figure\,\ref{orbitspec} that the iron K line at about 6.7\,keV is well identified for this data set. That feature in not so obvious in other spectra of the time series, certainly explaining the difficulty to fit the hardest thermal component. For the same reason as given in the case of HD\,168112, no non-thermal emission component was used to investigate any potential X-ray emission of inverse Compton scattering origin.

EPIC spectra obtained at the three epochs are shown in Fig.\,\ref{orbitspec}. The wide orbit of the triple system calculated on the basis of the ephemeris of \citet{Ibanoglu2013} is shown, along with the corresponding orbital phases of the observations (see Sect.\,\ref{var2} for a discussion). The dashed lines represent the stellar separation between components A an B, emphasizing the significant changes between 2002 and 2014.

\begin{table}
\caption{Best-fit parameters for the September 2014 EPIC-pn data set of HD\,167971 using a three-temperature model ({\tt wabs*wind*(apec+apec+apec)}, in the 0.3--10\,keV energy band). The error bars represent the 90\,\% confidence level.\label{3T167971}}
\begin{center}
\begin{tabular}{l l}
\hline
\vspace*{-0.2cm}\\
Parameter & Value \\
\hline
\vspace*{-0.2cm}\\
N$_{wind}$  &  0.45$_{-0.06}^{+0.05}$\,$\times$\,10$^{22}$\,cm$^{-2}$ \\
\vspace*{-0.2cm}\\
kT$_1$ &  0.29$_{-0.02}^{+0.02}$\,keV \\
\vspace*{-0.2cm}\\
N$_1$ &  8.26$_{-1.90}^{+2.37}$\,$\times$\,10$^{-3}$\,cm$^{-5}$ \\
\vspace*{-0.2cm}\\
kT$_2$ &  0.94$_{-0.05}^{+0.05}$\,keV \\
\vspace*{-0.2cm}\\
N$_2$ &  1.67$_{-0.26}^{+0.29}$\,$\times$\,10$^{-3}$\,cm$^{-5}$ \\
\vspace*{-0.2cm}\\
kT$_3$ & 1.98$_{-0.28}^{+0.61}$\,keV \\
\vspace*{-0.2cm}\\
N$_3$ &  0.66$_{-0.24}^{+0.21}$\,$\times$\,10$^{-3}$\,cm$^{-5}$ \\
\vspace*{-0.2cm}\\
\hline
\vspace*{-0.2cm}\\
$\chi^2_\nu$ (d.o.f.) & 0.95 (400) \\
f$_{obs}$ & 1.68$_{-0.07}^{+0.02}$\,$\times$\,10$^{-12}$\,erg\,cm$^{-2}$\,s$^{-1}$ \\
\vspace*{-0.2cm}\\
\hline
\end{tabular}
\end{center}
\end{table}

\subsection{Discussion}\label{disc2}

\subsubsection{The X-ray luminosity}\label{lumin2}
Considering the spectral types given by \citet{Ibanoglu2013} for the components of HD\,167971, typical bolometric luminosities can be estimated on the basis of the calibration by \citet{Martins2005}. These values, collected in Table\,\ref{lumin}, lead to a bolometric luminosity of 1.9 -- 2.6\,$\times$\,10$^{39}$\,erg\,s$^{-1}$. On the other hand, the X-ray flux (in the 0.3--10.0\,keV range) corrected for interstellar absorption can be estimated on the basis of the best-fit parameters, i.e. $\sim$\,10$^{-11}$\,erg\,cm$^{-2}$\,s$^{-1}$. For a distance to NGC\,6604 of 1700\,pc, one obtains an X-ray luminosity of about 3.5\,$\times$\,10$^{33}$\,erg\,s$^{-1}$. As a result, one obtains a L$_X$/L$_{bol}$ ratio of 1.3 -- 1.8\,$\times$\,10$^{-6}$, clearly pointing to a significant overluminosity certainly due to the colliding-wind regions. 

\begin{figure*}
\begin{center}
\includegraphics[width=16cm]{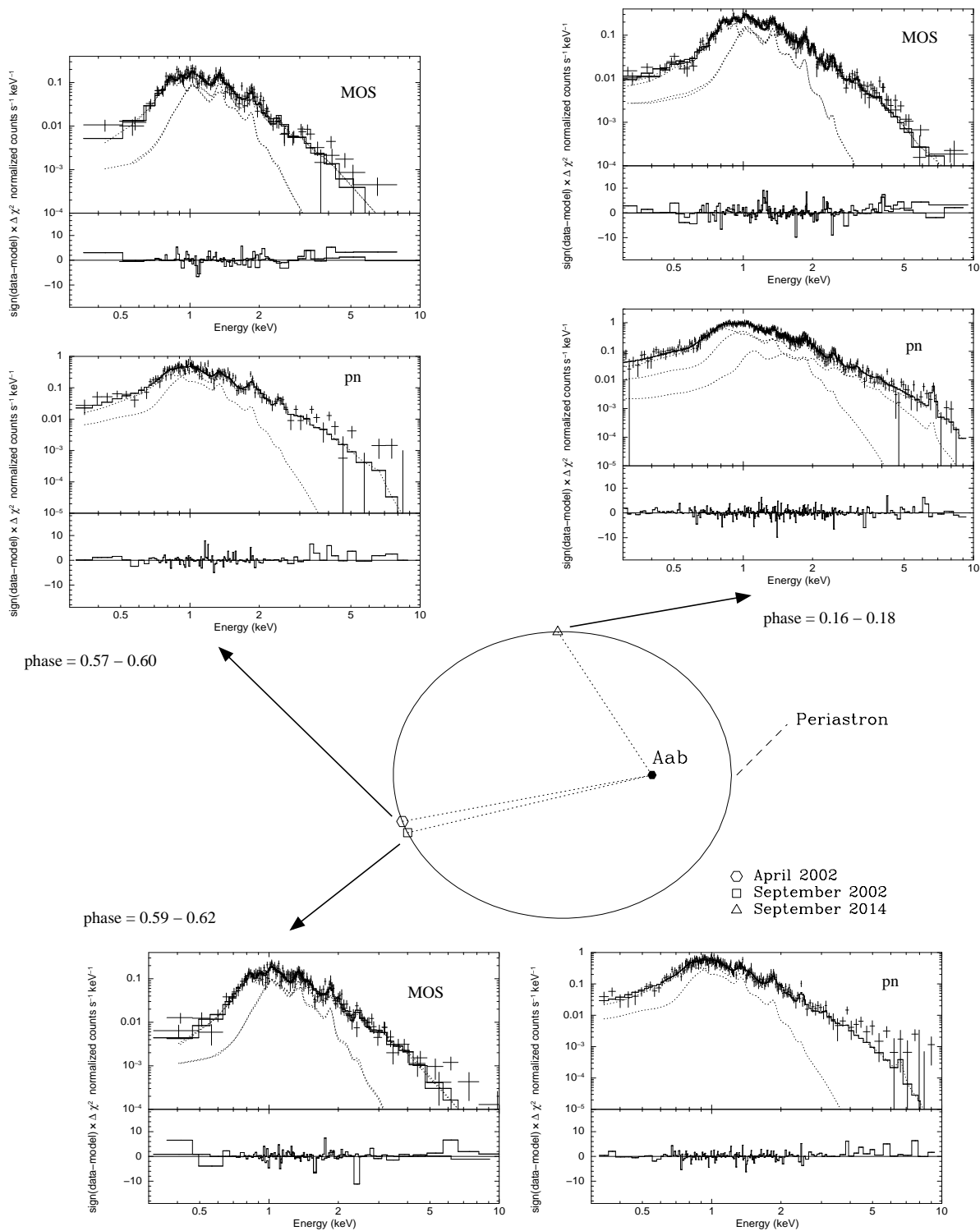}
\caption{Wide orbit of HD\,167971 based on the parameters published by \citet{Ibanoglu2013}. The plane of the page/screen coincides with the orbital plane. Periastron passage corresponds to an orbital phase equal to zero and dotted straight lines represent the stellar separation at the three epochs. Symbols (triangle, square and hexagon) represent the position of component B on the orbit at the three epochs (component Aab is at the focus on the right). The dark arrows point to the corresponding EPIC (MOS and pn) spectra along with their best-fit model. Individual model components are represented by the dotted curves. All models are two-temperature models, except for the pn spectra in September 2014 which is significantly best represented by a three-temperature model. \label{orbitspec}}
\end{center}
\end{figure*}

\begin{table}
\caption{Adopted parameters for the components in HD\,167971. The two values for the total bolometric luminosity come from the two hypotheses for the luminosity class of component B. \label{lumin}}
\begin{center}
\begin{tabular}{l c c c}
\hline
Comp. & Sp. type & L$_{bol}$  \\
  &  & (erg\,s$^{-1}$) \\
\hline
\vspace*{-0.2cm}\\
Aa & O7.5III & 8.9\,$\times$\,10$^{38}$ \\
Ab & O9.5III & 5.1\,$\times$\,10$^{38}$ \\
B  & O9.5III & 5.1\,$\times$\,10$^{38}$ \\
   & O9.5I   & 12.1\,$\times$\,10$^{38}$ \\
Total &  & 19.1\,$\times$\,10$^{38}$ \\
 &  & 26.1\,$\times$\,10$^{38}$ \\
\vspace*{-0.2cm}\\
\hline
\end{tabular}
\end{center}
\end{table}

On the basis of the wind parameters mentioned in Section\,\ref{system2} for components Aa and Ab, one can estimate a total kinetic power for component Aab of about 1.8\,$\times$\,10$^{36}$\,erg\,s$^{-1}$. For component B, assuming $\mathrm{{\dot M}_{B} = 1.4-3\,\times\,10^{-7}\,M_\odot\,yr^{-1}}$ and $\mathrm{V_{\infty,B}}$ = 2600-2900\,km\,s$^{-1}$ depending on the luminosity class (III or I), we obtain P$_{kin}$ $\sim$ 3 -- 8\,$\times$\,10$^{36}$\,erg\,s$^{-1}$. The total kinetic power in the triple system is thus $\sim$ 2.1--2.6\,$\times$\,10$^{36}$\,erg\,s$^{-1}$. The expected intrinsic emission from the individual stellar winds must be about 10$^{-7}$ times the bolometric luminosities. On the basis of the values quoted in Table\,\ref{lumin}, the expected intrinsic X-ray luminosity is in the range 1.9--2.6\,$\times$\,10$^{32}$\,erg\,s$^{-1}$. Subtracting this contribution from the total measured X-ray luminosity corrected for interstellar absorption, the contribution from the colliding-wind region is about 3.2-3.3\,$\times$\,10$^{33}$\,erg\,s$^{-1}$. The L$_X$/P$_{kin}$ ratio is thus in the range 1.2--1.6\,$\times$\,10$^{-3}$. It appears that the conversion rate of kinetic power into thermal soft X-ray emission from the colliding winds is an order of magnitude larger than for HD\,168112. This can be explained by the fact that two colliding-wind regions are present, and in particular one of them occurs in a close system in the radiative regime, highly efficient at producing X-rays. However, it is not possible with the present information to disentangle contributions of the X-ray luminosity coming from CWab and CWAB.

As a comparison, L$_X$/P$_{kin}$ ratio can also be estimated in the cases of two O + O PACWBs, one rather short period and one long period system. As a short period system, let us consider Cyg\,OB2\,\#8A (O6If + O5.5III(f)) whose orbital period is about 22 days \citep{Let8a}. The X-ray luminosity measured by \citet{DeBeckerxmm8a} was about 1.0-1.5\,$\times$\,10$^{34}$\,erg\,s$^{-1}$ (overluminous by a factor 20-30). For these spectral types, the estimated kinetic power (based on parameters by \citealt{Muijres2012}) is about 8.5\,$\times$\,10$^{36}$\,erg\,s$^{-1}$. This leads to a L$_X$/P$_{kin}$ ratio of about 1.2--1.8\,$\times$\,10$^{-3}$. For a long period system, the case of HD\,167974 might be relevant. It consists of an  O3.5V + O5-5.5V system, with a period of about 8.6\,yr \citep{rauw9sgr2012}. \citet{rauw9sgr} measured an X-ray luminosity of 1.4\,$\times$\,10$^{33}$\,erg\,s$^{-1}$. The spectral types of the components suggest a bolometric luminosity of about 3.4\,$\times$\,10$^{39}$\,erg\,s$^{-1}$. Removing the expected contributions from individual winds (10$^{-7}$ times L$_{bol}$), the X-ray contribution from the colliding winds is about 1.1\,$\times$\,10$^{33}$\,erg\,s$^{-1}$. These number allow to determine a L$_X$/P$_{kin}$ ratio of about 1.0\,$\times$\,10$^{-4}$. Among these two examples, HD\,167971 is closer to the case of Cyg\,OB2\,\#8A characterized by a much high conversion rate of kinetic power to thermal X-rays produced by the colliding winds. This emphasizes further the role played by the CWab region in the thermal X-ray spectrum of HD\,167971.

\subsubsection{Variability}\label{var2}

On the basis of the ephemeris published by \citet{Ibanoglu2013}, one can compute approximate orbital phases of the long orbit for the three epochs of XMM-Newton observations. Considering the large uncertainties on the orbital parameters, it is more realistic to give ranges of values, rather than unique values for the orbital phase. These ranges are 0.57--0.60, 0.59--0.62 and 0.16--0.18, respectively for A02, S02 and S14. With an eccentricity of about 0.53, the relative separation between components A and B is expected to drop by a factor 1/3 between 2002 and 2014 (see Fig.\,\ref{orbitspec} for a plot of the orbit). As a result, the thermal X-ray emission from CWAB is expected to be higher in 2014. However, it must be emphasized that the total thermal X-ray emission measured in the EPIC spectra comes from 5 contributions: (i) the wind of component Aa, (ii) the wind of components Ab, (iii) the wind of component B, (iv) the wind interaction CWab, and finally (v) the wind interaction CWAB. Among these contributions, only the last one is likely to change as a function of the orbital phase of the wide orbit. The anticipated change in the total measured X-ray flux should therefore be diluted. The measured fluxes do not point to any significant change of the X-ray flux between April 2002 and September 2014. 

The latter result appears quite surprising considering the X-ray brightness of the colliding-wind regions in HD\,167971 suggested by the L$_X$/L$_{bol}$ ratio. As a tentative explanation, one should note that an important fraction of the X-ray overluminosity may come from CWab, with only a moderate/minor contribution from CWAB. The short period (about 3.32\,d) in the Aab pair favours indeed a significantly high emission measure in CWab, in addition to the fact that the radiative regime is more efficient at producing thermal X-rays \citep[see e.g.][]{SBP1992,antokhin2004}. Considering the short separation in the eclipsing binary, the stellar winds have no chance to have reached their terminal velocity (see Section\,\ref{system2}). The post-shock temperature is therefore lower than in wider systems. The thermal X-ray emission from CWab is therefore expected to be rather soft. The fact that the global EPIC spectrum of HD\,167971 is not so hard supports the idea that the harder component expected to come from CWAB (with higher pre-shock velocities) does not dominate the X-ray emission. The plasma temperatures reported in Table\,\ref{fit167971TT} are indeed compatible with the pre-shock velocities estimated at the end of Section\,\ref{system2} for the Aab pair. If an important fraction of the X-ray overluminosity was attributed to CWAB, the lack of significant variation between 2002 and 2014 would suggest the observations actually occurred at somewhat symmetrical orbital phases, characterized by similar separations (and accordingly similar X-ray emission). This would however not be in agreement with the estimated orbital phases, and it is very unlikely that the uncertainty on the ephemeris of the wide orbit could lead to such a poor estimate of the orbital phases. Alternatively, this result may point to a lack of 1/D dependence of the the X-ray emission from CWAB, even though it would be at odd with the adiabatic nature of the shocked winds in the wide orbit. It is also interesting to note that the bright synchrotron radio emission along the wide orbit \citep{Blomme167971} provides compelling evidence that the shocks of the colliding-winds are clearly present. Otherwise, no particle acceleration could be operating. A wind-wind interaction region does not therefore need to be bright in X-rays to be efficient at accelerating particles. This lends additional support to the idea that particle acceleration -- and associated non-thermal radiative processes -- can efficiently operate even in a wide range of conditions.

Finally, now that approximate orbital elements are available, the marginal change of the X-ray flux between epochs A02 and S02 could certainly not be attributed to a change in the wide orbit. It could rather come from an eclipse-like event in the Aab close pair. On the basis of the revised ephemeris given by \citet{Ibanoglu2013}, the orbital phases in the Aab pair are 0.497, 0.922 and 0.446, respectively for epoch A02, S02 and S14. These numbers suggest indeed that the configuration of the system in September 2002 was likely to lead to a partial eclipse by the primary stellar wind, although in April 2002 and September 2014 the system was closer to an eclipse by the secondary. However, considering the short period of the Aab pair and the time elapsed since the determination of its ephemeris, the uncertainty on the corresponding orbital phase may be too large to definitely ascertain this interpretation.

\section{Summary and conclusions}\label{concl}
The X-ray emission from HD\,168112 and HD\,167971, two massive stellar systems in the open cluster NGC\,6604, has been investigated on the basis of archive and new data obtained with the XMM-Newton satellite. Thanks to updated information about their multiplicity revealed in the past last years, the results of the observations could be discussed and interpreted in a more appropriate context.

The X-ray spectrum of HD\,168112 appears to be slightly overluminous, in agreement with the idea that a fraction of the emission comes from the colliding-wind region. It is moderately hard, with higher plasma temperature of 15-30\,$\times$\,10$^{6}$\,K. The analysis of the present X-ray time series allowed to estimate a conservative lower limit on the eccentricity of about 0.38. The latter estimate is based on the hypothesis of adiabatic shocked winds, which seems to be validated by a careful inspection of wind parameters along with recent estimates of the projected separation obtained using long baseline interferometry. 

The X-ray spectrum of HD\,167971 is slightly softer than that of HD\,168112. Thanks to the existence of an orbital solution, the X-ray spectral time series could be interpreted with a good idea of the orbital phases of the observations. Considering the ephemeris published on the wide orbit, the marginal variability measured between April and September 2002 -- if real -- is more probably attributable to a partial wind eclipse in the short period Aab pair. The lack of significant variability between 2002 and 2014, despite the very different orbital phases, is certainly explained by a rather weak contribution of the X-ray emission coming from the colliding-wind region in the wide orbit (CWAB). The latter contribution is likely not bright enough to dominate the individual stellar contributions and that coming from the close eclipsing binary (CWab). This is in agreement with the radiative regime that prevails for the shocked winds in the close binary, allowing for a high efficiency in the conversion of the kinetic power of the winds into thermal X-ray emission. This efficiency is indeed expected to be significantly lower in the colliding-wind region in the wide orbit characterized by an adiabatic regime. This demonstrates that in hierarchical triple O-type systems such as HD\,167971, thermal X-rays do not necessarily constitute an efficient tracer to detect the long period wind-wind interaction. Interesting enough, a wind-wind interaction region does not need to be bright in X-rays to be efficient at accelerating particles. This lends additional support to the idea that particle acceleration can efficiently operate in a wide range of conditions.

This study allowed to achieve a better view of the X-ray emission from these two systems. The present measurements will be of great importance when they will be confronted to theoretical models aiming at simulating the physics of colliding-winds in massive binaries. In particular, these measurements are important for energy budget considerations that will have to be addressed in the context of the investigation of particle acceleration in massive binaries. Non-thermal processes -- including particle acceleration and radiative mechanisms -- occur on top the hydrodynamics of colliding-winds which is notably probed using X-ray measurements such as those discussed in the present study. 

From the observational point of view, campaigns are still in progress to derive the orbital elements of these systems using the VLTI. On the other hand, hard X-ray investigations are envisaged to probe the non-thermal X-ray emission from HD\,168112 and HD\,167971. As significant non-thermal radio emitters, the issue of the detection of non-thermal X-rays is still open. In this context, the information compiled in this study is intended to be used to prepare an overall X-ray investigation, including a soft thermal part (initiated in this study) and a hard non-thermal contribution likely to be probed with NuStar or even ASTRO-H (expected to be launched in late 2015).

\section*{Acknowledgments}

The author wants to warmly thank people working at the XMM-SOC for the scheduling of the XMM-Newton observations, along with Dr. Eric Gosset for interesting discussions and the anonymous referee for many useful comments that helped to significantly improve the paper. The SIMBAD database has been consulted for the bibliography. 

\bibliographystyle{mn2e}

%\bibliography{ngc6604}

\bsp

\label{lastpage}

\end{document}